\documentclass[12pt]{article}
\usepackage{geometry,amsmath,amssymb,epsfig}
\newcommand{\Iota}{\mathrm{I}}
\newcommand{\emdash}{---}
\newcommand{\mathe}{\mathrm{e}}
\newcommand{\tmem}[1]{{\em #1\/}}
\newcommand{\tmop}[1]{\ensuremath{\operatorname{#1}}}
\newcommand{\tmtextbf}[1]{{\bfseries{#1}}}
\newcommand{\tmtexttt}[1]{{\ttfamily{#1}}}

\newcommand{\Sh}[1]{#1\hskip-11pt \diagup}

\begin{document}

\title{Simulating the All-Order Hopping Expansion II: Wilson Fermions}
\author{
Ulli Wolff\thanks{
e-mail: uwolff@physik.hu-berlin.de} \\
Institut f\"ur Physik, Humboldt Universit\"at\\ 
Newtonstr. 15 \\ 
12489 Berlin, Germany
}
\date{}
\maketitle

\begin{abstract}
  We investigate the extension of the Prokof'ev-Svistunov worm algorithm to
  Wilson lattice fermions in an external scalar field. We effectively simulate
  by Monte Carlo the graphs contributing to the hopping expansion of the
  two-point function on a finite lattice to arbitrary order. Tests are
  conducted for a constant background field i.~e. free fermions at some mass.
  For the method introduced here this is expected to be a representative case.
  Its advantage is that we know the exact answers and can thus make stringent
  tests on the numerics. The approach is formulated in both two and three
  space-time dimensions. In $D = 2$ Wilson fermions enjoy special positivity
  properties and the simulation is similarly efficient as in the Ising model.
  In $D = 3$ the method also works at sufficiently large mass, but there is a
  hard sign problem in the present formulation hindering us to take the
  continuum limit.
\end{abstract}
\begin{flushright} HU-EP-08/61 \end{flushright}
\begin{flushright} SFB/CCP-08-98 \end{flushright}
\thispagestyle{empty}
\newpage

\section{Introduction}

To give a definition of quantum field theories beyond perturbation theory
(Feynman diagrams) they normally have to be regularized by replacing space and
usually also time by a lattice. Then the functional integral becomes a
well-defined object and thus amenable to numerical methods, usually in the
form of stochastically sampling lattice field configurations by Monte Carlo
methods. Most theories of interest contain fermion fields which lead to some
or all integration variables being anticommuting Grassmann `numbers'. In the
standard approach these integrations {\emdash} possibly after introducing
additional Bose fields {\emdash} are Gaussian and are performed exactly. The
result is an effective action of the bosonic fields alone which become coupled
non-locally. The known Monte Carlo techniques to practically simulate such
systems are mostly based on molecular dynamics and the hybrid Monte Carlo idea
(HMC) {\cite{Gottlieb:1987mq}}, {\cite{Duane:1987de}}. These methods have been
improved and optimized rather successfully over the years by a very large
effort of many members of the lattice community. On the other hand
practitioners know that once fermions are decoupled then HMC for locally
coupled Bose fields is not a very efficient algorithm compared to alternatives
like over-relaxation which are then available, not to mention the
(unfortunately few) cases where cluster methods can be applied. This implies a
large penalty for fermions even in cases where their effects are only small.
At small fermion masses the fermionic forces in HMC tend to grow and the step
size of the molecular dynamics trajectories has to be taken small enough. With
this quasi-continuous evolution one then has to be cautious about possible
long autocorrelations. After all, to the best of the author's knowledge, the
ergodicity of HMC has not been formally proven.

Maybe for the aforementioned reasons among others some part of the community
has remained motivated to look for radically different approaches. A rather
natural idea is to look for a representation of fermions as some sort of `sum
over configurations' more similar to the bosons. One of the pioneering papers
developing such ideas is {\cite{hirsch1982mcs}}. There as in numerous
succeeding attempts one starts from an operator formulation of fermions and
inserts intermediate states in the occupation number basis between factors of
the transfer matrix. In this way occupied sites map out an ensemble of
`world-lines' or a gas of loops of fermions on the lattice. Often the
amplitudes that arise oscillate in sign with the danger of leading to an
unmanageable signal to noise ratio, the infamous fermionic sign problem. The
inclusion of gauge fields in this approach poses additional problems.

A somewhat different approach was successful for {\emdash} but also restricted
to {\emdash} QCD at infinite gauge coupling, $\beta = 0$,
{\cite{Rossi:1984cv}}. In the Euclidean path integral with staggered fermions
but no gauge plaquette term the group valued gauge fields can be integrated
out first. The resulting model of locally paired even Grassmann elements has
contributions that can be viewed as a statistical system of explicitly
color-neutral mesons (dimers) and baryon loops. Both these systems and the
world-line gas \`a la {\cite{hirsch1982mcs}} are difficult to simulate
efficiently by local methods due to constraints which conflict with local
deformations of the configurations. In some cases efficient nonlocal updates
could be devised {\cite{Evertz:2000rk}}, {\cite{Adams:2003cca}}.

A `more Euclidean' version of the idea was proposed in
{\cite{Karowski:1984ih}}. These authors started from the determinant of the
integrated-out staggered fermions and tried to stochastically generate its
expansion into cycles. However the restriction to local updates and the
sign problem, even for $D = 2$ in this case, have limited the use of the
method.

In {\cite{Gattringer:1998cd}} (see also {\cite{Scharnhorst:1996gj}}) a loop or
world-line representation was proposed for the partition function of standard
two-dimensional Euclidean Wilson fermions in an external scalar field. They
were mapped on a certain 8-vertex model. Based on it the Gross-Neveu model was
simulated with local updates in {\cite{Gattringer:2007em}}. In
{\cite{Wolff:2007ip}} the same representation was re-derived directly from the
Grassmann integral for charge self-conjugate (Majorana) Wilson fermions. The
mapping between Wilson fermions and a loop-gas could in addition be made
precise also for a finite torus with (anti)periodic boundary conditions. A
cluster algorithm for the loop-gas was developed in {\cite{Wolff:2007ip}}
which produces almost uncorrelated loop configurations at low cost. In the
sequel Willi Rath and the author have tried to compute correlations based on
these configurations {\cite{WRdip}}. The only solution we have found so-far
proceeds via the numerical generation of the scalar $\sigma$-field that
usually factorizes the Gross-Neveu interaction. Then the close to singular
Dirac operator in this random scalar field has to be inverted and the CPU time
ends up being spent in a very similar fashion as in HMC.

In this paper, as an alternative approach, we adopt the `worm' algorithm of
Prokof'ev and Svistunov (PS) {\cite{prokofev2001wacci}} to lattice fermions of
the Wilson type. To this end we build on the study of the PS algorithm for the
Ising model carried out as a preparation in {\cite{Wolff:2008km}}. While there
the (untruncated) strong coupling expansion is sampled, the fermion loop-gas
corresponds to the quite similar hopping expansion{\footnote{Because of this
strong similarity, we put this paper into one series with
{\cite{Wolff:2008km}}.}}. We here extend the loop-gas formulation of fermions
on a torus in two ways. We generalize {\cite{Wolff:2007ip}} to including two
spinor field insertions at arbitrary lattice sites. It turns out that the PS
algorithm is ideally suited to keep track of the non-local amplitudes involved
due to Fermi statistics. The second non-trivial extension takes this
construction to Majorana fermions in three Euclidean dimensions. While we
understand why the fermionic sign problem mentioned before is absent in two
dimensions if the system size is large in correlation lengths, the full
problem has to be confronted in three dimensions. We indeed find for free
fermions that are implemented numerically in this study, that the PS algorithm
for $D = 2$ is similarly efficient as in the Ising model. While clearly
correct in principle also in $D = 3$ it fails numerically with the present
technique when the continuum limit is approached. We nonetheless find the
three dimensional loop representation theoretically quite interesting. We
think that the free Majorana fermion in $D = 3$ is an excellent study ground
for more clever techniques, for instance cluster improved observables, to
still overcome the sign problem, perhaps along the lines of
{\cite{Chandrasekharan:1999cm}}.

The organization of this paper is as follows. In the next section we set up
our notation for the lattice fermions discussed followed by section 3
introducing dimers that label all possible hopping graphs needed for the PS
simulation. Tools for the simulations are described in 4. In section 5. we define the
kind of observables on the loop ensemble that allow to make contact with
fermionic two-point functions followed in 6. by the description of numerical
results. We end on 7. conclusions including a brief outline how interaction
can be added. In two appendices we collect the free fermion results used as
benchmarks and a geometrical discussion of the fermionic phase factors arising
for each closed loop.

\section{Majorana-Wilson lattice fermions}

We start from a standard Wilson-Dirac fermion with the action
\begin{equation}
  S_{\tmop{WD}} = a^D \sum_x \overline{\psi} (\gamma_{\mu}
  \tilde{\partial}_{\mu} + m - \frac{r}{2} a \partial^{\ast} \partial) \psi .
  \label{SDirac}
\end{equation}
We consider a $D$-dimensional standard hypercubic lattice with spacing $a$ in
all directions and either periodic or antiperiodic boundary conditions for
each direction over the respective periodicity length $L_{\mu}$. The boundary
conditions are coded into a vector $\varepsilon_{\mu}$ with components 0,1 by
the condition
\begin{equation}
  \psi (x \pm L_{\mu}  \hat{\mu}) = (- 1)^{\varepsilon_{\mu}} \psi (x)
\end{equation}
and similarly for $\overline{\psi}$, and $\hat{\mu}$ is a unit vector in the
positive \ $\mu$ direction.

Unless stated otherwise, the mass $m$ is assumed to be a real $x$-dependent
periodic external field $m (x)$ here. By later integrating over it with a
suitable weight one can, starting from this building block, arrive at
interacting theories like the Gross-Neveu model. The operators $\partial,
\partial^{\ast}, \tilde{\partial}$ are the usual forward, backward, and
symmetrized nearest neighbor differences. The set $\{\gamma_{\mu}, \mu = 0, 1,
\ldots, D - 1\}$ are hermitean Euclidean Dirac matrices. From here on \ we
shall restrict ourselves to the space-time dimensions $D = 2, 3$ with $2
\times 2$ $\gamma$-matrices in both cases. The Wilson term suppresses the
doublers and from here on we set its coefficient to the convenient value $r =
1$.

The action (\ref{SDirac}) is invariant under charge conjugation for any $m
(x)$. It is hence both possible and natural to split the fermion into two
neutral Majorana components by setting
\begin{equation}
  \psi = \frac{1}{\sqrt{2}} (\xi_1 + i \xi_2), \hspace{1em} \overline{\psi} =
  \frac{1}{\sqrt{2}} (\xi_1^{\top} - i \xi_2^{\top})\mathcal{C}
\end{equation}
with the charge conjugation matrix $\mathcal{C}$ obeying
\begin{equation}
  \text{$\mathcal{C}$} \gamma_{\mu}  \text{$\mathcal{C}$}^{- 1} = -
  \gamma_{\mu}^{\top} = - \gamma_{\mu}^{\ast}, \quad \text{$\mathcal{C}$} = -
  \text{$\mathcal{C}$}^{\top} .
\end{equation}
Inserting this into (\ref{SDirac}) we find two identical contributions for
$\xi_{1, 2}$. In our Majorana reduction we consider only one such component in
the following
\begin{equation}
  S = \frac{1}{2} a^D \sum_x \xi^{\top} \mathcal{C}(\gamma_{\mu}
  \tilde{\partial}_{\mu} + m - \frac{1}{2} a \partial^{\ast} \partial) \xi .
\end{equation}
Note that the matrix in this quadratic form is antisymmetric. By collecting
diagonal and neighbor terms we can rewrite this action as
\begin{equation}
  S = \frac{1}{2} \sum_x (D + m) \xi^{\top}_{} (x)\mathcal{C} \xi (x) -
  \sum_{x, \mu} \xi^{\top}_{} (x)\mathcal{C}P ( \hat{\mu}) \xi (x + \hat{\mu})
  \label{Majo}
\end{equation}
where we now have adopted lattice units ($a = 1$) and have introduced
projectors
\begin{equation}
  P (n) = \frac{1}{2} (1 - n_{\mu} \gamma_{\mu}) \hspace{1em} (n^2 = 1)
\end{equation}
for each lattice direction ($n = \pm \hat{\mu})$. Note that the hopping term
of a Majorana fermion is a function of the {\tmem{unoriented}} link because of
the identity
\begin{equation}
  \xi^{\top}_{} (x)\mathcal{C}P ( \hat{\mu}) \xi (x + \hat{\mu}) =
  \xi^{\top}_{} (x + \hat{\mu})\mathcal{C}P (- \hat{\mu}) \xi (x) .
  \label{nonorient}
\end{equation}
For $D = 2$ the form (\ref{Majo}) coincides with the starting point of
{\cite{Wolff:2007ip}}.

To continue we introduce the shorthand notation
\begin{equation}
  \overline{\xi} = \xi^{\top} \mathcal{C}.
\end{equation}
We emphasize that for the Majorana fermion this depends on $\xi$ while $\psi,
\overline{\psi}$ were independent Grassmann integration variables. The
partition function is given by
\begin{equation}
  Z_0^{^{(\varepsilon)}} = \int D \xi \mathe^{- S}  = \tmop{Pf}
  [\mathcal{C}(\gamma_{\mu} \tilde{\partial}_{\mu} + m - \frac{1}{2}
  \partial^{\ast} \partial)] \label{Z0eps}
\end{equation}
where the Gaussian integral over Majorana fields has led to a
Pfaffian{\footnote{The order of factors in $D \xi$ is assumed to be such that
this is true without an (irrelevant) extra sign.}} of the antisymmetric
matrix. The result depends on the boundary conditions, of course, which is
exhibited for $Z_0^{(\varepsilon)}$ but left implicit on the right hand side.

In a straight-forward generalization of {\cite{Wolff:2008km}} we now extend
our study to include
\begin{equation}
  Z^{(\varepsilon)} (u, v) = \int D \xi \mathe^{- S}  \xi (u) \overline{\xi}
  (v), \label{Zuv}
\end{equation}
which is a matrix in spin space. It is closely related to the two point
function{\footnote{The dependence of $G$ on the boundary conditions
$\varepsilon$ is left implicit.}}
\begin{equation}
  G (x, y ; m) = \langle \xi (x) \overline{\xi} (y) \rangle =
  \frac{Z^{^{(\varepsilon)}} (x, y)}{Z_0^{^{(\varepsilon)}}} .
\end{equation}
As we are considering bilinear fermions in an external field $m (x)$ the
propagator can also be obtained as the solution of a system of linear
equations
\begin{equation}
  (\gamma_{\mu} \tilde{\partial}_{\mu} + m - \frac{1}{2} \partial^{\ast}
  \partial) G (x, y ; m) = \delta_{x, y} \times 1_{\tmop{spin}} \label{Geq}
\end{equation}
where the Dirac operator acts on $x$.

For constant $m$ such an evaluation can proceed by Fourier expansion and will
serve us as a check below. Otherwise the Pfaffian is a problem similar to the
fermion determinant and methods like HMC are suitable at least for an even
number of flavors {\cite{Korzec:2006hy}}, {\cite{TomPhD}}. Our objective here
is however to develop a simulation method alternative to this approach.

We end this section with the remark that, in contrast to the Ising model,
$Z^{(\varepsilon)} (x, x)$ is not equal to the ordinary partition function.
Instead one may show that for any $m$
\begin{eqnarray}
  Z^{(\varepsilon)} (x, x) = \frac{\partial Z_0^{^{(\varepsilon)}}}{\partial m
  (x)} & \times 1_{\tmop{spin}} \label{Sdensity} & 
\end{eqnarray}
holds. To derive this relation we use that the space of antisymmetric $2
\times 2$ matrices is only one-dimensional, given by multiples of \ the second
Pauli matrix. Hence in $Z^{(\varepsilon)} (x, x)$ the integral containing \
$\xi (x) \xi (x)^{\top}$ must be proportional to $\mathcal{C}^{- 1}$.

\section{Dimer form of Majorana fermions}

We here derive the loop-gas form of the fermion correlation function and
partition function. In principle this may be achieved by using theorems for
the expansion of the Pfaffian together with the sparseness of the matrix
introduced before. This would parallel the approach in
{\cite{Karowski:1984ih}}. Instead we shall extensively manipulate the
representation by a Grassmann integral. This is physically more transparent
and may be seen as deriving the required expansion formulas `on the fly' as
they are needed. The power of Grassmann numbers for such purposes was
emphasized before in {\cite{Samuel:1978zx}}.

\subsection{General structure}

We start from the factorized form
\begin{equation}
  Z^{(\varepsilon)} (u, v) = \int D \xi \prod_z  \mathe^{- \frac{1}{2} \varphi
  (z) \overline{\xi} (z) \xi (z)}  \left[ \prod_{l = \langle x y \rangle} 
  \mathe^{\overline{\xi} (x) P ( \widehat{y - x}) \xi (y)}  \right] \xi (u)
  \overline{\xi} (v)
\end{equation}
with the short hand
\begin{equation}
  \varphi (x) = D + m (x) .
\end{equation}
Because each $P$ is a one-dimensional projector and due to the Grassmann
nature of $\xi$ there are only two terms{\footnote{While many of the previous
steps go through also for $D = 4$, we will need a third term here.}} in the
expansion of each link-factor. It can thus be `dimerized'
\begin{equation}
  \mathe^{\overline{\xi} (x) P ( \widehat{y - x}) \xi (y)} = \sum_{k_l = 0,
  1} [ \overline{\xi} (x) P ( \widehat{y - x}) \xi (y)]^{k_l}
\end{equation}
leading to
\begin{equation}
  Z^{(\varepsilon)} (u, v) = \sum_{\{k_l \}} \int D \xi \prod_z  \mathe^{-
  \frac{1}{2} \varphi \overline{\xi}^{} \xi}  \left[ \prod_{l = \langle x y
  \rangle} [ \overline{\xi} (x) P ( \widehat{y - x}) \xi (y)]^{k_l} \right]
  \xi (u) \overline{\xi} (v) . \label{Zuvk}
\end{equation}
For each configuration $\{k_l \}$ we say that links with $k_l = 1$ carry an
(active) dimer. Associated with each site we have only two Grassmann variables
integrated over. This implies numerous constraints on contributing dimer
configurations:
\begin{itemize}
  \item at sites $x \notin \{u, v\}$ there can only be either 0 or 2 dimers
  adjacent
  
  \item if $u \not= v$, at these two sites there must be exactly \ 1 dimer
  touching
  
  \item at $u = v$ there can be no dimer touching.
\end{itemize}
In the simpler case of $Z_0^{^{(\varepsilon)}}$ the analogous expansion
requires 0 or 2 dimers around all sites. Any dimer configuration that obeys
these conditions and contributes to $Z^{(\varepsilon)} (u, v)$ or to
$Z_0^{^{(\varepsilon)}}$ we call {\tmem{admissible}}.

As a consequence of these constraints dimers in admissible configurations have
to form chains. These can never backtrack, intersect or overlap. For \ $u
\not= v$ there must be exactly one chain or string connecting $u$ and $v$
that we call $\sigma$. Apart from it all other chains must form a number of
closed loops $\lambda_j$. For contributions to \ $Z_0^{^{(\varepsilon)}}$
there is no string but only loops and the same is true for $Z^{(\varepsilon)}
(u, u)$ with the additional requirement that no loop passes through $u$.

It is to be emphasized that the string (if present) and the loops including
their number is a unique one-to-one representation of an admissible dimer
configuration,
\begin{equation}
  \text{} \{k_l \} |_{\tmop{admissable}} \longleftrightarrow \sigma \cup
  \{\lambda_j, j = 1, 2, \ldots, N_{\lambda} \},
\end{equation}
where both the string and the set of loops can also be empty.

\subsection{Amplitudes}

To evaluate a contribution {\tmem{for fixed admissible }}$\{k_l \}$, we can
reorder freely all Grassmann bilinears like hopping terms of links with $k_l =
1$, the local $\varphi \overline{\xi}^{} \xi$ terms and the integration
measure with two spin components $d \xi_1 d \xi_2$ for each site. In this way
the whole Grassmann integral is factorized into one site integrals that are
carried out with the formula
\begin{equation}
  \int \text{$d \xi_1 d \xi_2$} \xi \overline{\xi} = 1_{\tmop{spin}},
\end{equation}
where we have chosen $\mathcal{C}_{12} = 1 = -\mathcal{C}_{21}$. At each
monomer site {\emdash} a site with no dimer adjacent {\emdash} the
integrations are saturated by a site-factor in (\ref{Zuvk}) and contribute a
factor $\varphi (z) = D + m (z)$.

Next we consider $\sigma$ and define $| \sigma |$ to be the number of dimers
one has to cross to walk from $u$ to $v$. During the walk one encounters a
sequence of sites $s_i$ separated by lattice unit vectors $n_i$,
\begin{equation}
  \tmop{string} \sigma \leftrightarrow \{ \text{$u = s_0, s_1, s_2, \ldots,
  s_{| \sigma |} = v$} \}, \hspace{1em} n_{i + 1} = s_{i + 1} - s_i .
\end{equation}
We notice that we may use (\ref{nonorient}) to our convenience along the path.
Then, after carrying out the integrations belonging to all sites $s_i$, there
emerges a product of projectors
\begin{equation}
  V (\sigma) = P (n_1) P (n_2) \cdots P (n_{| \sigma |}) .
\end{equation}
Each closed loop $\lambda_j$ can be labeled in exactly the same way except
that now $s_0 = s_{| \lambda |}$ holds (suppressing for the moment the loop
index $j$). If we denote the sequence of unit shifts now by $\{m_1^{}, m_2,
\ldots, m_{| \lambda |} \}$ , then closed loops contribute a scalar factor
\begin{equation}
  w (\lambda_j) = - \tmop{tr} [P (m_1) P (m_2) \cdots P (m_{| \lambda |})] .
  \label{wlam}
\end{equation}
The minus sign here is the usual one coming from closed fermion loops.
Technically speaking, upon closing the trace, one pair of $\xi,
\overline{\xi}$ appears in the `wrong' order. The cyclicity of the trace
immediately implies, that $w (\lambda_j)$ is independent of where we start
with $m_1$ along the loop. In addition one may use
\begin{equation}
  P (n)^{\top} =\mathcal{C} P (- n) \mathcal{C}^{- 1}
\end{equation}
to show also independence of the direction chosen to traverse the loop. Hence
$w (\lambda_j)$ is truly a function of the unoriented loop only.

\subsection{Evaluation of spin factors\label{spinfac}}

Using a bra-ket notation in spin space we write the Wilson projectors as
\begin{equation}
  \left. \left. P (m) = | m \right\rangle \langle m |, \quad \langle m | m
  \right\rangle = 1, \hspace{1em} m = \pm \hat{\mu} = \pm \hat{0}, \pm \hat{1}
  [, \pm \hat{2} \tmop{if} D = 3] . \label{eigenspinor}
\end{equation}
Now the loop factor is composed of scalar products
\begin{equation}
  \left. \left. \left. \left. w (\lambda_j) = - \langle m_1 | m_2
  \right\rangle \langle m_2 | m_3 \right\rangle \cdots \langle m_{| \lambda |
  - 1} | m_{| \lambda |} \right\rangle \langle m_{| \lambda |} | m_1
  \right\rangle,
\end{equation}
i.e. factors associated with the sites connecting pairs of links met along the
loop. The modulus of the individual factors is one where successive $m_i$
coincide (straight sections) and $1 / \sqrt{2} = \cos (\pi / 4)$ between
orthogonal links (corners). This is easily seen from a simple example
\begin{equation}
  | \langle \hat{0} | \hat{1} \rangle |^2 = \langle \hat{0} |P ( \hat{1}) |
  \hat{0} \rangle = \langle \hat{0} | \gamma_0 P ( \hat{1}) \gamma_0 | \hat{0}
  \rangle = \langle \hat{0} |P (- \hat{1}) | \hat{0} \rangle = \frac{1}{2}
\end{equation}
and similarly for any other orthogonal pair. With $\pi / 4$ we see the typical
half-angle appearing with spinors. We thus find
\begin{equation}
  w (\lambda_j) = 2^{- C (\lambda_j) / 2} \phi (\lambda_j),
\end{equation}
where $C (\lambda_j)$ is the number of corners around the loop and $\phi
(\lambda_j)$ is a phase.

We now discuss a rather direct way to compute $\phi (\lambda_j)$. While it
gives not much geometric insight into its meaning, this derivation will lend
itself to a very direct algorithmic implementation. In appendix \ref{appB} an
alternative more geometrical analysis is presented.

We fix the ambiguous phases of $| \pm \hat{\mu} \rangle$ in a definite way,
knowing that $w$ is independent of this convention. We start from $| \hat{0}
\rangle$ with an arbitrary phase. Next one may demand a maximal number of five
real positive phases
\begin{equation}
  \langle \pm \hat{1} | \hat{0} \rangle = \langle - \hat{0} | \hat{1} \rangle
  = \langle \pm \hat{2} | \hat{0} \rangle = \frac{1}{\sqrt{2}} .
\end{equation}
This exhausts the free choices and the remaining phases, one in $D = 2$ and
additional six in $D = 3$ can be {\tmem{evaluated}}. One possible way to do so
is to construct all eigenvectors starting from $| \hat{0} \rangle$ with the
help of the projectors:
\begin{equation}
  \text{$| \pm \hat{1} \rangle$=} \sqrt{2} P (\pm \hat{1}) | \hat{0} \rangle,
\end{equation}
\begin{equation}
  \text{$| - \hat{0} \rangle$=$\sqrt{2} P (- \hat{0})$} | \hat{1} \rangle = -
  \gamma_1 | \hat{0} \rangle
\end{equation}
and
\begin{equation}
  \text{$| \pm \hat{2} \rangle$=} \sqrt{2} P (\pm \hat{2}) | \hat{0} \rangle .
\end{equation}
\begin{table}[htb]
\centering
    \begin{tabular}{|c|c|c|c|c|c|c|}
      \hline
      & $| + \hat{0} \rangle$ & $| - \hat{0} \rangle$ & $| + \hat{1} \rangle$
      & $| - \hat{1} \rangle$ & $| + \hat{2} \rangle$ & $| - \hat{2}
      \rangle$\\
      \hline
      $\langle + \hat{0} |$ & 1 & - & 1 & 1 & 1 & 1\\
      \hline
      $\langle - \hat{0} |$ & - & 1 & $1$ & $- 1$ & $i$ & $- i$\\
      \hline
      $\langle + \hat{1} |$ & 1 & 1 & 1 & - & $z$ & $z^{\ast}$\\
      \hline
      $\langle - \hat{1} |$ & 1 & $- 1$ & - & 1 & $z^{\ast}$ & $z$\\
      \hline
      $\langle + \hat{2} |$ & 1 & $- i$ & $z^{\ast}$ & $z$ & 1 & -\\
      \hline
      $\langle - \hat{2} |$ & 1 & $i$ & $z$ & $z^{\ast}$ & - & 1\\
      \hline
    \end{tabular}
  
  \caption{Phases in all possible scalar products between
  eigenspinors.\label{tab1}}
\end{table}

{\noindent}The implied phases can now be computed by just using the Dirac
algebra and they are collected in table \ref{tab1}. In two dimensions only the
upper left $4 \times 4$ block is relevant, where all phases are real, they
belong to Z(2). In this case the sign for each loop acquires a simple
geometrical interpretation which will be discussed in section \ref{phases}.

In evaluating the phases involving the third dimension we have assumed that
$\gamma_1 \gamma_2 = - i \gamma_0$ holds and set
\begin{equation}
  z = \frac{1 + i}{\sqrt{2}} = \mathe^{i \frac{\pi}{4}} .
\end{equation}
We cannot eliminate the complex phase factors by re-defining the phases of
$\text{$| \pm \hat{2} \rangle$}$. There is another inequivalent irreducible
Dirac representation in $D = 3$ with the complex conjugate phases (for example
from $\gamma_{\mu} \rightarrow - \gamma_{\mu})$. This means that the parity
reflected loop has the opposite phase which is hence a pseudoscalar. The phase
factors for 3-dimensional Wilson fermions are in Z(8) and all values are
actually assumed for relatively short loops if non-planar ones are included.
The group Z(8) is related to the rotation group being reduced to the lattice
symmetries, see appendix \ref{appB}. Simple examples for loops with complex phases
are shown in figure \ref{cloop}. They were extracted from a Monte Carlo
simulation (see below) be `tagging' the phase of configurations and plotting
them.

\begin{figure}[htb]
  \centering
 \epsfig{file=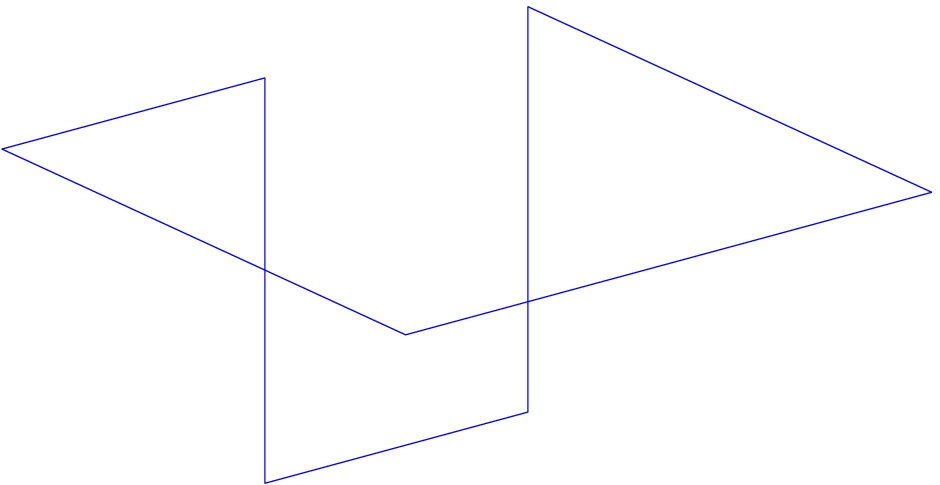,width=6.0cm}
 \qquad
 \epsfig{file=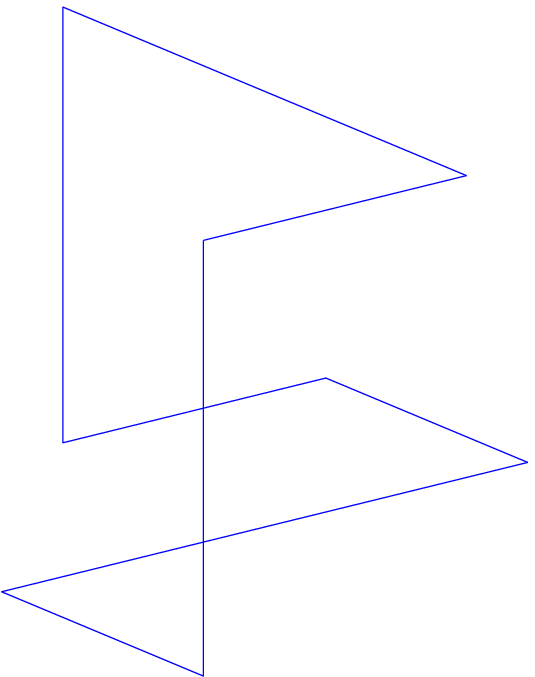,width=3.0cm}
  \caption{Closed fermion loops in $D = 3$ with phase $\exp (i \pi / 4)$
  (left) and $\exp (i \pi / 2)$ (right).\label{cloop}}
\end{figure}

The spin factor for the open string from $u$ to $v$ is given by
\begin{equation}
  \left. \left. V (\sigma) = |n_1 \rangle \langle n_1 | n_2 \right\rangle
  \cdots \langle n_{| \sigma | - 1} | n_{| \sigma |} \right\rangle \langle
  n_{| \sigma |} |.
\end{equation}
For the leftmost ket and the rightmost bra we introduce the notation
\begin{equation}
  |n_1 \rangle = |n (u) \rangle, \hspace{1em} \langle n_{| \sigma |} | =
  \langle n (v) | \hspace{1em} (u \not= v)
\end{equation}
such that $n (u)$ is the unit vector pointing {\tmem{out}} of $u$ in the
direction of the unique adjacent dimer $k_l = 1$ while $n (v)$ is the
corresponding unit vector pointing {\tmem{toward}} $v$. Note that in principle
we should write $n (u ; k)$ and $n (v ; k)$ and both vectors are undefined if
$u = v$ holds. The scalar factors have again a modulus $2^{- 1 / 2}$ for each
corner and a phase $\phi (\sigma)$ that may be constructed from table
\ref{tab1},
\begin{equation}
  V (\sigma) = 2^{- C (\sigma)} \phi (\sigma) \; |n (u) \rangle \langle n (v)
  |.
\end{equation}

We re-emphasize that all objects discussed above including the number of
corners, the string and loop decomposition and the various phases are
(nonlocal) functions of the $k_l$ in an admissible configuration. It would
however clutter our notation too much to always exhibit this explicitly.

\subsection{Boundary conditions\label{bcfac}}

More phase factors can arise from boundary conditions if loops or the string
winds around the torus in antiperiodic directions an odd number of times. We
adopt the convention to label the points on the torus by coordinates $x_{\mu}
= 0, 1, \ldots, L_{\mu} - 1$ and distinguish a $(D-1)$-dimensional sheet of
`boundary' links{\footnote{Of course, the torus has no boundary, hence the
quotes.}} for each direction as follows:
\begin{equation}
  l \tmop{is} a \tmop{boundary} \tmop{link} \tmop{in} \tmop{direction} \mu
  \leftrightarrow \text{$l = \langle x, x + \hat{\mu} \rangle$ with $x_{\mu} =
  L_{\mu} - 1$} .
\end{equation}
For the string $\sigma$ and for each loop $\lambda_j$ we introduce parities
$e_{\mu} (\sigma$) and $e_{\mu} (\lambda_j)$ defined by
\begin{equation}
  e_{\mu} (\sigma) = \left\{ \begin{array}{ll}
    1 & \tmop{if} \sigma \tmop{contains} \tmop{an} \tmop{odd} \tmop{number}
    \tmop{of} \mu - \tmop{boundary} \tmop{links}\\
    0 & \tmop{else}
  \end{array} \right. \label{blink}
\end{equation}
and for the closed loops $e_{\mu} (\lambda_j)$ is completely analogous. The
overall sign from the boundary conditions is now given by
\begin{equation}
  \tmop{sign} = (- 1)^{\varepsilon \cdot \overline{e}} \hspace{1em}
  \tmop{with} \hspace{1em} \overline{e}_{\mu} = e_{\mu} (\sigma) + \sum_{j =
  1}^{N_{\lambda}} e_{\mu} (\lambda_j) \hspace{1em} (\tmop{mod} 2)
\end{equation}
with the scalar product of the $D$-vectors $\varepsilon_{\mu}$ and
$\overline{e}_{\mu}$ in the exponent.

All contributions to the amplitude of an admissible dimer configuration have
now been identified and will be combined in the next section to represent
fermionic quantities in an ensemble summing and ultimately sampling such
configurations.

\section{Dimer partition function and worm algorithm}

\subsection{Dimer partition function}

Next we formally write down the characteristic function $\Theta (k ; u, v)$
that is unity for admissible configurations and zero for all others. In
principle it has been defined before in words. As a building block we use
\begin{equation}
  d (k ; x) = \sum_{l, \partial l \ni x} k_l
\end{equation}
which counts the number of dimers adjacent at $x$. Then we define
\begin{eqnarray}
  \Theta (k ; u, v) & = & \delta_{d (k ; u), 1} \delta_{d (k ; v), 1} 
  \prod_{x \not{\in} \{u, v\}}  \left( \delta_{d (k ; x), 0} + \delta_{d (k ;
  x), 2} \right) \hspace{1em} \tmop{for} \; u \not= v \\
  \Theta (k ; u, v) & = & \prod_x  \left( \delta_{d (k ; x), 0} + \delta_{d (k
  ; x), 2} \right) \hspace{1em} \tmop{for} \; u = v. 
\end{eqnarray}
Note that the constraint enforced for $u = v$ here is the one for
contributions to $Z_0^{(\varepsilon)}$ rather than the more restrictive one
for contributions to $Z^{(\varepsilon)} (u, u)$. We now consider the following
partition function
\begin{equation}
  \mathcal{Z}= \sum_{u, v, \{k_l \}} \frac{\Theta (k ; u, v)}{\rho^{} (u, v)}
  2^{- \overline{C} / 2}  \prod_{x, d (k ; x) = 0} \varphi (x) . \label{Zdim}
\end{equation}
Here $\overline{C}$ is the total number of corners
\begin{equation}
  \overline{C} = C (\sigma) + \sum_{j = 1}^{N_{\lambda}} C (\lambda_j),
\end{equation}
$\rho$ is an arbitrary symmetric strictly positive lattice-periodic function
similar as in {\cite{Wolff:2008km}}. The product is the weight from all
monomer sites and we here restrict ourselves to
\begin{equation}
  \varphi (x) = D - \text{$m (x) > 0$}
\end{equation}
guaranteeing the positivity of the overall weight. Expectation values of
observables $A (k ; u, v)$ in this ensemble are defined by
\begin{equation}
  \left\langle \left\langle A \right\rangle \right\rangle =
  \frac{1}{\mathcal{Z}}  \sum_{u, v, \{k_l \}} A (k ; u, v) \frac{\Theta (k ;
  u, v)}{\rho^{} (u, v)} 2^{- \overline{C} / 2}  \prod_{x, d (k ; x) = 0}
  \varphi (x) . \label{primobs}
\end{equation}

Observables related to the Majorana fermions discussed before can be \ written
as ratios of such expectation values. This will be discussed in the next
section after introducing the simulation algorithm for (\ref{Zdim}).

\subsection{Prokof'ev-Svistunov worm algorithm}

The simulation of the dimer ensemble can be carried out with the worm
algorithm of PS {\cite{prokofev2001wacci}}. It is very similar to the
algorithm described in {\cite{Wolff:2008km}} and we can be brief here about
details. The main difference to the Ising case is that more than 2 dimers
cannot touch and that there is a weight $1 / \sqrt{2}$ for corners which
induces a kind of stiffness (tendency to be straight) of the chains.

We briefly pause here to comment on the notation of hopping parameter
expansion \ in the title of the paper. The factors $\varphi (x)$ appearing for
monomers could be rescaled to unity by absorbing them into $\xi (x)$ early on.
Then each dimer $k_{\langle x y \rangle}$ would be accompanied by a factor
$[\varphi (x) \varphi (y)]^{- 1 / 2}$. For constant $m$ this would equal $[D +
m]^{- 1} = 2 \kappa$ with the conventional hopping parameter $\kappa$. Thus \
$2 \kappa$ is the strict analog of the Ising strong coupling expansion
parameter $\tan \beta$ in {\cite{Wolff:2008km}}. We prefer however to stay
with the unrescaled form which is advantageous for the introduction of
interaction via $m (x)$.

\begin{figure}[htb]
  \centering

  
 \epsfig{file=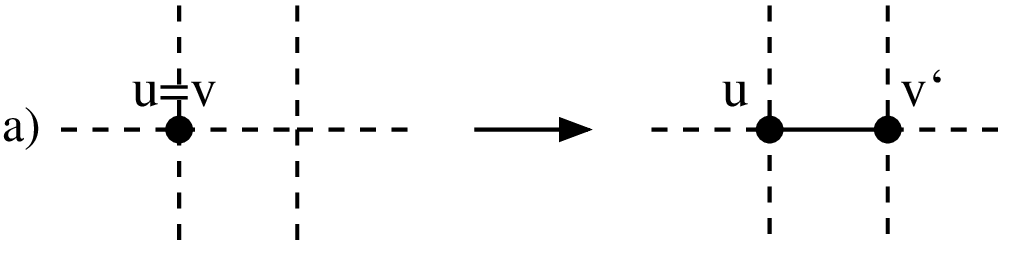,width=0.6\textwidth}\\[1ex]
 \epsfig{file=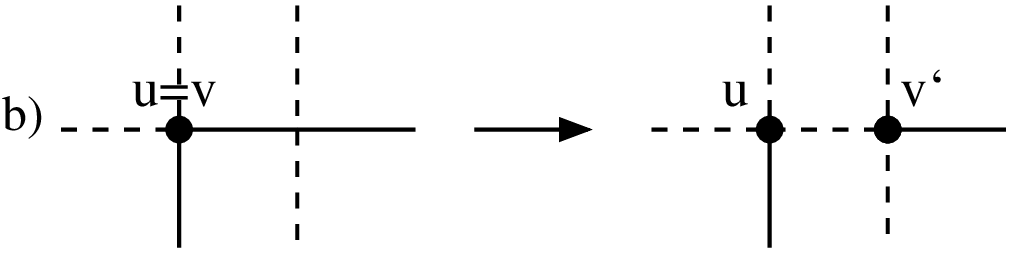,width=0.6\textwidth}\\[1ex]
 \epsfig{file=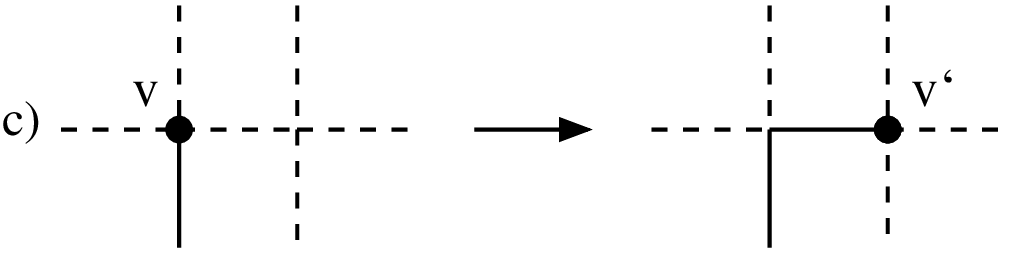,width=0.6\textwidth}

  \caption{Pictorial representation of elementary moves in the PS algorithm.
  Solid lines are active dimers (value one).\label{movefig}}
\end{figure}

An update microstep of the PS algorithm is now a succession of steps I and II
applied to admissible configurations. In step I we make a Metropolis decision
on a proposal where we pick one of the $2 D$ nearest neighbors of $v$ with
equal probability and call it $v'$ and the connecting link $l$. The proposed
move changes $v \rightarrow v'$ \ flipping at the same time $k_l \rightarrow
k'_l = 1 - k_l$. It brings us from the global configuration $k$ to
configuration $k'$ (differing at exactly one link). Note that $k'$ may be not
admissible, in which case the move will be rejected. We first form an
auxiliary quantity $q$, the ratio of amplitudes after and before the move. We
have to distinguish a number of cases and collect values of $q$ in table
\ref{tabq}. We have written $n (u')$ as a shorthand for what should be $n (u ;
k')$ {\tmem{after}} the move, although we did not move $u$ here. The allowed
moves are illustrated in figure \ref{movefig}. Parts a), b), c) refer to lines
1, 3, 5 of the table. The lines below those refer to the reverse changes. They
correspond to the same graphs read from right to left with the arrow reversed
and interchanged $v \leftrightarrow v'$. The directions of active dimers not
participating in the present update are examples only and can also point
differently.

\begin{table}[htb]
\centering
    \begin{tabular}{|c|c|c|c|c|}
      \hline
      $d (k ; v)$ & $d (k ; v')$ & $d (k', v)$ & $d (k' ; v')$ & $q$\\
      \hline
      0 & 0 & 1 & 1 & $[\varphi (v) \varphi (v')]^{- 1}$\\
      \hline
      1 & 1 & 0 & 0 & $\varphi (v) \varphi (v')$\\
      \hline
      2 & 2 & 1 & 1 & $- [\langle n (v') |v - v' \rangle \langle v - v' |n
      (u') \rangle]^{- 1}$\\
      \hline
      1 & 1 & 2 & 2 & $- \langle n (v) |v' - v \rangle \langle v' - v|n (u)
      \rangle$\\
      \hline
      1 & 0 & 2 & 1 & $\langle n (v) |v' - v \rangle [\varphi (v')]^{- 1}$\\
      \hline
      1 & 2 & 0 & 1 & $\varphi (v) [\langle n (v') |v - v' \rangle]^{- 1}$\\
      \hline
    \end{tabular}
  \caption{Entries in the first four columns specify the condition for
  possible moves, under which the amplitude gets multiplied by $q$ (not including
  possible signs from antiperiodic boundary conditions).\label{tabq}}
\end{table}

{\noindent}In all other cases not covered here $q$ is set to zero$.$ In lines
3 and 4 of the table a sign is included for changing the number of fermion
loops $N_{\lambda}$ by one. Finally the modulus of $q$ is used in the
acceptance probability
\begin{equation}
  p_{\tmop{acc}} = \min \left( 1, \frac{\rho (u, v)}{\rho (u, v')} | q|
  \right)
\end{equation}
while the phase changes will be considered in the next section.

The type II move is as follows. If we encounter a configuration $u, v, \{k_l
\}$ with $u = v$ we `kick' $u = v$ together to a randomly chosen other lattice
site with unchanged $\{k_l \}$ with the probability $0 < p_{\tmop{kick}}
\leqslant 1$. For $u \not= v$ we do nothing in this step, which is the
dominant case. A difference with respect to the Ising case is that while there
also $p_{\tmop{kick}} = 0$ (absence of step II) yields an ergodic algorithm
{\cite{Deng:2007jq}}, this is not so here. For the fermions the jumps are
required to move between different connected components.

Moving only $v$ together with steps II constitutes a correct Monte Carlo
algorithm. We nevertheless found it advantageous to also move $u$ in a
completely analogous fashion. We thus now call the sequence $\Iota_u -
\tmop{II} - \Iota_v - \tmop{II}$ a microstep and call $N_x$/2 mircosteps an
iteration if we have $N_x$ lattice sites.

We found the choice of $p_{\tmop{kick}} \in [0.3, 1]$ not critical and use
$p_{\tmop{kick}} = 0.7$ in the following after a few quick experiments.

\section{Fermionic phase and spin factors\label{phases}}

\subsection{Formulae for both $D = 2$ and $D = 3$}

The configurations of the dimer ensemble just discussed correspond to the set
of graphs of the hopping parameter expansion of Majorana fermions. Each
admissible configuration contributes to $Z^{(\varepsilon)}_0$ or to
$Z^{(\varepsilon)} (u, v)$ with a certain amplitude and in the second case
also with a spin matrix. The moduli of the amplitudes have been incorporated
into the generation of configurations. The phases of the amplitudes and the
spin matrices will be taken into account now as observables evaluated as in
(\ref{primobs}).

We first combine all phases discussed in section \ref{spinfac} and \ref{bcfac}
into the total phase
\begin{equation}
  \Phi^{(\varepsilon)} (k) = (- 1)^{\varepsilon \cdot \overline{e}} \phi
  (\sigma) \prod_{j = 1}^{N_{\lambda}} \phi (\lambda_j) .
\end{equation}
For $u = v$ we set $\phi (\sigma) = 1$, there is no string, only loops.

As mentioned before for $D = 2$ the phase $\Phi$ is just a sign while for $D =
3$ it is an element of Z(8). We now have the connection
\begin{equation}
  V Z^{(\varepsilon)}_0 =\mathcal{Z} \left\langle \left\langle \rho (u, u)
  \delta_{u, v} \Phi^{(\varepsilon)} (k) \right\rangle \right\rangle
  \label{Z0dim}
\end{equation}
with the volume
\begin{equation}
  V = \prod_{\mu = 0}^{D - 1} L_{\mu} .
\end{equation}
If we define a spin matrix
\begin{equation}
  \mathcal{S}(k ; u, v) = \left\{ \begin{array}{lll}
    |n (u) \rangle \langle n (v) | & \tmop{if} & u \not= v\\
    \delta_{d (k ; u), 0} [\varphi (u)]^{- 1} 1_{\tmop{spin}} & \tmop{if} & u
    = v
  \end{array} \right.
\end{equation}
the cases with insertions may be uniformly written as
\begin{equation}
  Z^{(\varepsilon)} (x, y) = \rho (x, y)\mathcal{Z} \left\langle \left\langle
  \delta_{u, x} \delta_{v, y} \Phi^{(\varepsilon)} (k)\mathcal{S}(k ; u, v)
  \right\rangle \right\rangle . \label{Zxy}
\end{equation}
At coinciding arguments $x = y$ the Grassmann integrations are saturated by
the insertion alone, which requires a monomer site with its usual weight
factor to be canceled. In this case one could in principle also relax the
constraint in (\ref{Zxy}) to obtain
\begin{equation}
  V Z^{(\varepsilon)} (x, x) =\mathcal{Z} \left\langle \left\langle \rho (u,
  u) \delta_{u, v} \Phi^{(\varepsilon)} (k)\mathcal{S}(k ; x, x) \right\rangle
  \right\rangle .
\end{equation}
For the fermion correlation function the connection is
\begin{equation}
  G (x, y ; m) = \frac{\rho (x, y) \left\langle \left\langle \delta_{u, x}
  \delta_{v, y} \Phi^{(\varepsilon)} (k)\mathcal{S}(k ; u, v) \right\rangle
  \right\rangle}{(1 / V) \left\langle \left\langle \rho (u, u) \delta_{u, v}
  \Phi^{(\varepsilon)} (k) \right\rangle \right\rangle} . \label{Gxy}
\end{equation}
An alternative derivation of the result for coinciding points starts from the
observation
\begin{equation}
  - \frac{1}{2} \left\langle \overline{\xi} \xi (x) \right\rangle =
  \frac{\partial}{\partial m (x)} \ln Z^{(\varepsilon)}_0 = \frac{1}{\varphi
  (x)}  \frac{\left\langle \left\langle \rho (u, u) \delta_{u, v}
  \Phi^{(\varepsilon)} (k) \delta_{d (k ; x), 0} \right\rangle
  \right\rangle}{\left\langle \left\langle \rho (u, u) \delta_{u, v}
  \Phi^{(\varepsilon)} (k) \right\rangle \right\rangle}
\end{equation}
and uses (\ref{Z0dim}) and (\ref{Sdensity}).

From the structure of the contributions in the dimer ensemble we may conclude
that the right hand side of (\ref{Gxy}) is rational in the external field $m
(x)$. The denominator has total degree $V$, the numerator $V - 1 - | \sigma
|_{\min}$. Here $| \sigma |_{\min}$ is the minimal number of links to connect
$x$ and $y$ by a string. The degree in each individual $m (x)$ is only linear
both in the numerator and denominator.

The above formulae simplify if translation invariance holds, $m (x) \equiv m$,
$G (x, y ; m)$ $ \rightarrow G (x - y)$, where we also restrict $\rho (x, y) =
\rho (x - y)$ and normalize $\rho (0) = 1$. We then find
\begin{equation}
  G (z) = \rho (z) \frac{ \left\langle \left\langle \delta^{(\varepsilon)}_{u
  - v, z} \Phi^{(\varepsilon)} (k)\mathcal{S}(k ; u, v) \right\rangle
  \right\rangle}{\left\langle \left\langle \delta_{u, v} \Phi^{(\varepsilon)}
  (k) \right\rangle \right\rangle} .
\end{equation}
We recognize close similarities with the Ising correlation in
{\cite{Wolff:2008km}} with the novelty of averaging the phasefactor and the
spin matrices. Note that the delta function $\delta^{(\varepsilon)}$ in the
numerator needs to have the same antiperiodicity as the fields $\xi$.

From (\ref{Z0dim}) one may now trivially read off that
\begin{equation}
  \frac{Z^{(\varepsilon)}_0}{Z^{(\varepsilon')}_0} = \frac{\left\langle
  \left\langle \delta_{u, v} \Phi^{(\varepsilon)} (k) \right\rangle
  \right\rangle}{\left\langle \left\langle \delta_{u, v} \Phi^{(\varepsilon')}
  (k) \right\rangle \right\rangle} \label{Zrat}
\end{equation}
allows to measure the change in free energy for different boundary conditions.

This type of quantity is theoretically nice, because it is expected to
possess a continuum limit in a finite volume. For the massless all-periodic
case $\varepsilon_{\mu} \equiv 0$, the partition function
$Z_0^{(\varepsilon)}$ vanishes at $m = 0$ because the matrix under the
Pfaffian then has two exact zero modes. The corresponding phasefactor then
averages to zero exactly.

\subsection{$D = 2$ specialties}

Fermions in two Euclidean or one space dimension are simpler and in a way
untypical for the true problem related to the oscillating phase $\Phi$. In the
Euclidean field theory formulation this is seen by the phases from fermion
loops and from spin `essentially canceling' in $D = 2$ (only). In our
realization this is seen as follows. Minus signs appear only at two types of
corners, namely $\langle - \hat{0} | - \hat{1} \rangle$ or $\langle - \hat{1}
| - \hat{0} \rangle$. By drawing closed loops with the intersection properties
relevant here on a planar torus it is not difficult so see that
\begin{itemize}
  \item loops winding around the torus in one or both dimensions receive an
  even number of such minus signs
  
  \item loops that close trivially and do not wind around the torus receive an
  odd number of minus signs from spin phases.
\end{itemize}
In {\cite{Wolff:2007ip}} a more detailed discussion of this and some
illuminating figures with examples can be found. Winding around the torus can
be read off from the crossing of `boundary links' (\ref{blink}). Thus the result
for each closed loop on the two-dimensional torus can be summarized in our
notation as
\begin{equation}
  \phi (\lambda_j) = \left\{ \begin{array}{lll}
    + 1 & \tmop{if} & e_{\mu} (\lambda_j) = (0, 0)\\
    - 1 & \tmop{else} & 
  \end{array} \right. \hspace{1em} (D = 2 \tmop{only}) .
\end{equation}
Negative signs {\emdash} remember that the Fermi loop sign has been included
in $\phi$ {\emdash} only come from topologically nontrivial loops that cannot
be contracted to the trivial loop by series of plaquette moves
{\cite{Gattringer:2007em}}, {\cite{Wolff:2007ip}}. The total phase for $u = v$
configurations can now be given as
\begin{equation}
  \Phi^{(\varepsilon)} (k) = (- 1)^{\varepsilon \cdot \overline{e} +
  \delta_{\overline{e}, (0, 0)} + 1} \hspace{1em} [\tmop{for} u = v, \Theta (k
  ; u, u) = 1] . \label{Phi2D}
\end{equation}
It depends on $k$ {\tmem{only via}} the topology variable
$\overline{e}_{\mu}$. Using Fourier transformation on Z(2)
[$\sum_{\varepsilon} (- 1)^{\varepsilon \cdot (e - e')} = 4 \delta_{e, e'}$]
one may show the identity
\begin{equation}
  1 \equiv \sum_{\varepsilon} z (\varepsilon) \Phi^{(\varepsilon)} (k),
  \hspace{1em} z (\varepsilon) = \frac{1}{2}  (- 1)^{\delta_{\varepsilon, (0,
  0)}}
\end{equation}
for this case. This in turn implies for the average monomer density (with no
phases)
\begin{equation}
  \overline{K} = \frac{1}{V}  \frac{\left\langle \left\langle \delta_{u, v} 
  \sum_x \delta_{d (k ; x), 0} \right\rangle \right\rangle}{ \left\langle
  \left\langle \delta_{u, v} \right\rangle \right\rangle} \label{Kbar}
\end{equation}
the exact result (for free fermions)
\begin{equation}
  \overline{K} = \frac{2 + m}{V}  \frac{\partial}{\partial m} \ln
  \overline{Z}_0
\end{equation}
with the partition function
\begin{equation}
  \overline{Z}_0 = \sum_{\varepsilon} z (\varepsilon) Z^{(\varepsilon)}_0
\end{equation}
summed over boundary conditions with amplitudes $z (\varepsilon)$. Similarly
from (\ref{Zrat}) we may deduce now
\begin{equation}
  \frac{Z^{(\varepsilon)}_0}{\overline{Z}_0} = \frac{\left\langle \left\langle
  \delta_{u, v} \Phi^{(\varepsilon)} (k) \right\rangle
  \right\rangle}{\left\langle \left\langle \delta_{u, v} \right\rangle
  \right\rangle} . \label{Phibar}
\end{equation}

\section{Numerical experiments}

In this section we test our simulation method for the case of free fermions on
various lattice sizes and (constant) $m$ values in both $D = 2$ and 3. While
of course no numerical simulations are really needed here, we nonetheless
think that for our loop-gas representation this is not an untypical case also
for later interacting applications. Thus the advance knowledge of the results
here is just an advantage for precision testing.

To extract fermionic quantities from simulations we must have the phase
$\Phi^{(\varepsilon)} (k)$ available for each sampled configuration for the
desired boundary conditions. For each admissible configuration with $\Theta (k
; u, v) = 1$ it can be constructed by tracing the string and all loops at the
cost of order $V$ operations. It is however easier to update its value
together with the configurations. In fact this is even necessary to measure
efficiently between microsteps as discussed in {\cite{Wolff:2008km}}. To that
end we assume $\Phi^{(\varepsilon)} (k)$ to be known for the start
configuration. We always took the trivial $u = v = 0, k_l \equiv 0$ with
$\Phi^{(\varepsilon)} (k) = 1$. Then, whenever an \ update proposal of type
$\Iota_v$ of the PS algorithm is accepted, we change
\begin{equation}
  \Phi^{(\varepsilon)} (k) \rightarrow \Phi^{(\varepsilon)} (k') =
  \Phi^{(\varepsilon)} (k) \times \frac{q}{|q|} \times \eta (\langle v v'
  \rangle, \varepsilon)
\end{equation}
and similarly for $\Iota_u$. Here $q$ is given in table \ref{tabq} and the
additional factor
\begin{equation}
  \eta (\langle v v' \rangle, \varepsilon) = \left\{ \begin{array}{lll}
    - 1 & \tmop{if} & \langle v v' \rangle \tmop{is} \text{a} \mu -
    \tmop{boundary} \tmop{link} \tmop{and} \varepsilon_{\mu} = 1\\
    + 1 & \tmop{else} & 
  \end{array} \right.
\end{equation}
takes into account the boundary conditions (see (\ref{blink})). Needless to
say, one may also keep track of $\Phi^{(\varepsilon)}$ for several boundary
conditions in the same run, as the updates do not depend on them.

The spin matrix is easy to construct at any time from the single dimers
adjacent to $u$ and $v \not= u$ and is trivial for the coinciding case. In
practice we measure correlations contracted with some Dirac matrix $\Gamma$
which leads to
\begin{equation}
  - \left\langle \overline{\xi} (0) \Gamma \xi (z) \right\rangle = \rho (z)
  \frac{ \left\langle \left\langle \delta^{(\varepsilon)}_{u - v, z}
  \Phi^{(\varepsilon)} (k) \tmop{tr} [\mathcal{S}(k ; u, v) \Gamma]
  \right\rangle \right\rangle}{\left\langle \left\langle \delta_{u, v}
  \Phi^{(\varepsilon)} (k) \right\rangle \right\rangle} . \label{kGammara}
\end{equation}
We want to further specialize to zero spatial momentum as discussed in
appendix \ref{appA},
\begin{equation}
  k_{\Gamma} (z_0) = - \sum_{z_k} \left\langle \overline{\xi} (0) \Gamma \xi
  (z) \right\rangle \propto \rho (z_0) \left\langle \left\langle
  \delta^{(\varepsilon)}_{u_0 - v_0, z_0} \Phi^{(\varepsilon)} (k) \tmop{tr}
  [\mathcal{S}(k ; u, v) \Gamma] \right\rangle \right\rangle . \label{kGamma}
\end{equation}
We took $\rho$ to only depend on time and dropped the denominator. For
symmetry reasons only $\Gamma = 1, \gamma_0$ (also labeled as $S, V$) leading
to the scalar and vector correlations $k_S, k_V$ are nontrivial. During the
simulation we simply add the corresponding amplitudes into bins for each
separation $z_0$ and then end up with correlations measured for all distances.
Pre-computed tables are heavily used to speed up and they lead to a very
simple code. In the simulations reported below we have observed Metropolis
acceptance rates close to 50\% \ for $D = 2$ and $30\% \tmop{for} D = 3$. This
is close to the amplitude change by $1 / (D + m)$ when the worm `eats' a
monomer which it has to do to grow. In equilibrium also the other processes
are important, but this one seems to set the scale.

All error estimates below are derived with the method and tools detailed in
{\cite{Wolff:2003sm}}. In particular the definition of integrated
autocorrelation times $\tau_{\tmop{int}}$ employed here can be found, see also
remarks in {\cite{Wolff:2008km}}. Due to time series of length $10^6$ and more
the convolution step in \tmtexttt{UWerr}, eq. (31) in {\cite{Wolff:2003sm}},
became a bit slow for online data analysis. We therefore tailored a special
version \tmtexttt{UWerr\_fft} which accelerates this step by using the fast
Fourier transform. It is available on the web under
\tmtexttt{www.physik.hu-berlin.de/com/ALPHAsoft}.

\begin{figure}[htb]
  \centering
\epsfig{file=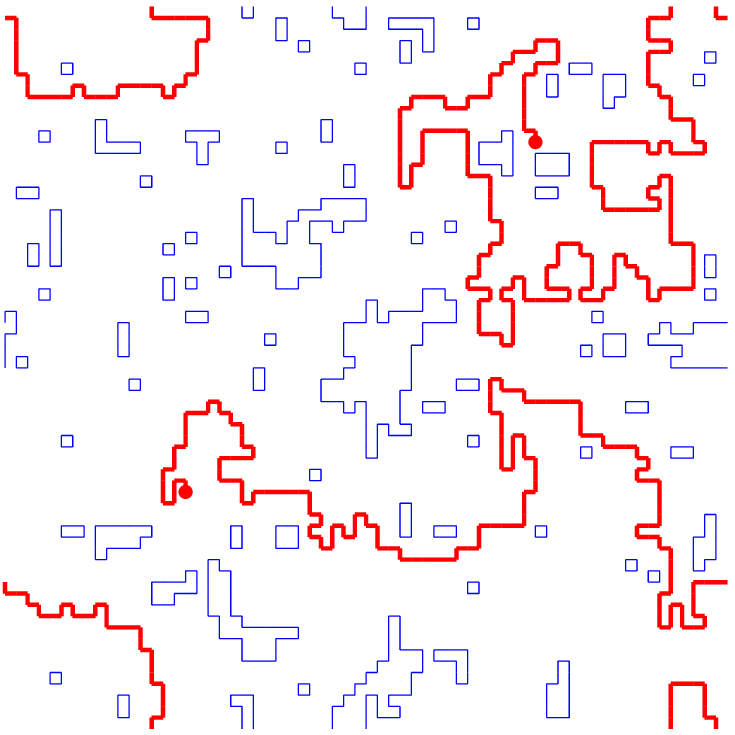,width=0.45\textwidth}
\qquad
\epsfig{file=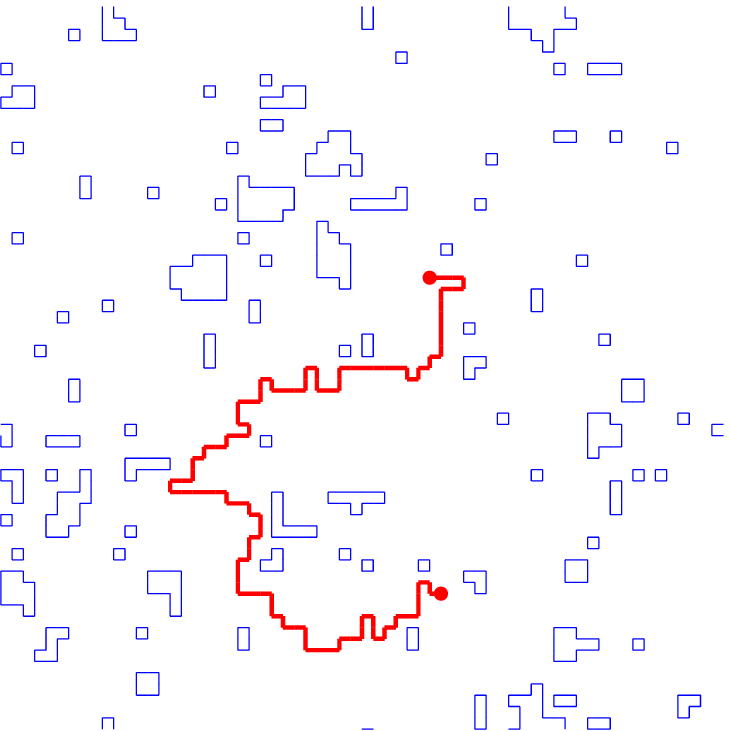,width=0.45\textwidth}

  \caption{Two typical configurations on a $64^2$ lattice at criticality ($m =
  0$, left panel) and with correlation length $64 / 5$ ($m = 0.0812 \ldots .$,
  right panel). The string $\sigma$ is given by the fat (red line), the other
  lines are loops $\lambda_j$. Readers are asked to identify left-right and
  top-bottom edges in their mind.\label{D2conf}}
\end{figure}

\subsection{$D = 2$, physically large volume}

For tests in this subsection we chose a mass such that $\omega L = 5$ holds
with the pole mass $\omega = \ln (1 + m)$. The zero momentum timeslice
correlations (\ref{kGamma}) then fall off exactly with $\exp (- \omega x_0$),
modified to cosh or sinh due to time periodicity, see appendix \ref{appA}. A
typical configuration is visualized by the right picture in figure
\ref{D2conf}. We see that for fermions, in contrast to the Ising model, there
really is a unique `worm', which moves by the updates{\footnote{The poor unoriented
Majorana worm has however no distinction between head and tail!}}. In all
simulations the complete zero momentum two-point functions at all separations
were consistent within errors with the exact results. We routinely checked
plots of the deviation in units of the estimated error against $x_0$ which are
order one, occasionally straddling $\pm 2$. \ In addition diagnostic
quantities like (\ref{Kbar}) and (\ref{Phibar}) were monitored. Because in a
large box (compared to the inverse mass) few configurations wind around the
torus we find no significant difference between periodic and antiperiodic
boundary conditions. In the example to follow we measured (\ref{Phibar}) and
obtained
\[ \frac{Z^{(0, 0)}_0}{\overline{Z}_0} = 0.9744 (5), \hspace{1em} \frac{Z^{(1,
   0)}_0}{\overline{Z}_0} = 0.9745 (5) \]
in agreement with the exact answer.

\begin{figure}[htb]
  \centering
\epsfig{file=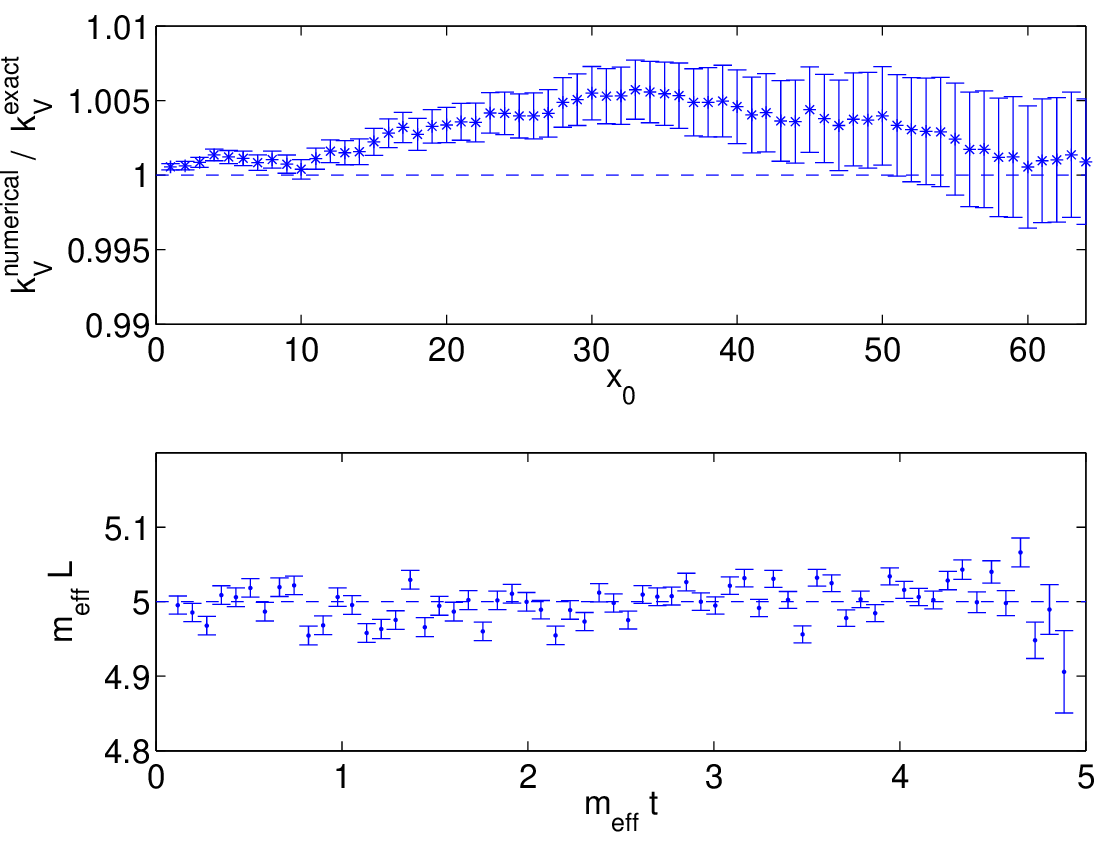,width=0.9\textwidth}
  \caption{Correlation function $k_V$ and the effective mass derived from it.
  Errorbars are one sigma high. \label{fig2}}
\end{figure}

As an impression for the reader we show in figure \ref{fig2} results for the
vector correlation $k_V$ on a lattice $L = 64, T = 2 L$ with
$\varepsilon_{\mu} = (1, 0)$ after accumulating $10^7$ iterations (steps per
site). The correlation length $\omega^{- 1}$ is hence about 13 lattice
spacings. We have chosen the bias $\rho$ as
\begin{equation}
  \rho (t) \propto \cosh [\omega (T / 2 - t)],
\end{equation}
which leads to a population of timeslices $\langle \langle \delta_{u_0 - v_0,
t} \rangle \rangle$ approximately flat in $t$. We refer the reader to the
discussion in {\cite{Wolff:2008km}}, which can be taken over essentially
without change. The upper panel shows the correlation $k_V (x_0)$ itself
normalized by its exact value. The growth of the errors from left to right is
due to an increase of the integrated correlation time $\tau_{\tmop{int}}$ from
about 0.6 iterations at short distances to about 15 iterations at $x_0 = T /
2$. In the lower panel we give the effective mass as a function of distance by
matching subsequent timeslices to
\begin{equation}
  \frac{k_V (x_0 + 1)}{k_V (x_0)} = \frac{\cosh (m_{\tmop{eff}} (T / 2 - x_0 -
  1)}{\cosh (m_{\tmop{eff}} (T / 2 - x_0)}, \hspace{1em} 0 < m_{\tmop{eff}}
  \equiv m_{\tmop{eff}} (x_0 + 1 / 2) . \label{meff}
\end{equation}
Here errors appear (apart from $x_0$ very close to $T / 2$) to be independent
of the separation in agreement with the observed autocorrelations
$\tau_{\tmop{int}} \approx 0.5$ for all $x_0$. The longer autocorrelations
observed in $k_V$ apparently cancel in the ratio. From the fluctuations in
figure \ref{fig2} we conclude qualitatively that statistical fluctuations at
neighboring time separations are strongly correlated in $k_V$, but much less
so in $m_{\tmop{eff}}$. In a run with $\rho \equiv 1$ the growth of
$\tau_{\tmop{int}}$ for $k_V$ does not occur. Its error however grows in a
similar way due to the larger variance coming from fewer data at large
separation $u - v$ (fewer `long worms') when no bias $\rho$ is applied. The
more interesting effective mass is more accurate with the bias used for the
figure.

\subsection{$D = 2$, physically small volume}

We now simulate at the critical point which in the free case is known to occur
at $m = 0.$ We remind that due to the infrared regulator given by the small
{\tmem{inverse}} temperature $T$ with antiperiodic boundary conditions
$\varepsilon_0 = 1$ this is physically well-defined. Such applications are of
interest in interacting theories to study nonperturbative renormalization
using the universal finite volume continuum limit. To this end we report
measurements of $k_S (T / 4)$ and $k_V (T / 2)$ with $\varepsilon = (1, 0)$.
Further motivation for the study of these objects can be found in appendix
\ref{appA} and refs. {\cite{Korzec:2006hy}}, {\cite{TomPhD}}. In table
\ref{tab3} we compile our results from performing $10^6$ iterations at each of
the lattice sizes. Again $\tau_{\tmop{int}}$ are given in iterations.

\begin{table}[htb]
\centering
  \begin{tabular}{|c|c|c|c|c|}
    \hline
    $T = L$ & $k_S (T / 4)$ & $\tau_{\tmop{int}, k_S (T / 4)}$ & $k_V (T / 2)$
    & $\tau_{\tmop{int}, k_V (T / 4)}$\\
    \hline
    16 & 0.010(7) & 1.85(8) & 0.997(10) & 1.11(2)\\
    \hline
    32 & -0.011(15) & 3.6(4) & 1.007(16) & 1.43(4)\\
    \hline
    64 & 0.011(25) & 5.0(5) & 0.943(29) & 2.68(14)\\
    \hline
  \end{tabular}
  \caption{Results at the critical point $m = 0$. The exact values for all $L$
  are $k_S (T / 4) = 0$ due to chiral symmetry and $k_V (T / 2) = 1$
  corresponding to canonical field normalization.\label{tab3}}
\end{table}

H{\noindent}ere the topology and the sign $\Phi^{(1, 0)}$ fluctuate, but we
can achieve a percent accuracy with the given statistics, which could be
enlarged. 

\begin{figure}[htb]
  \centering
\epsfig{file=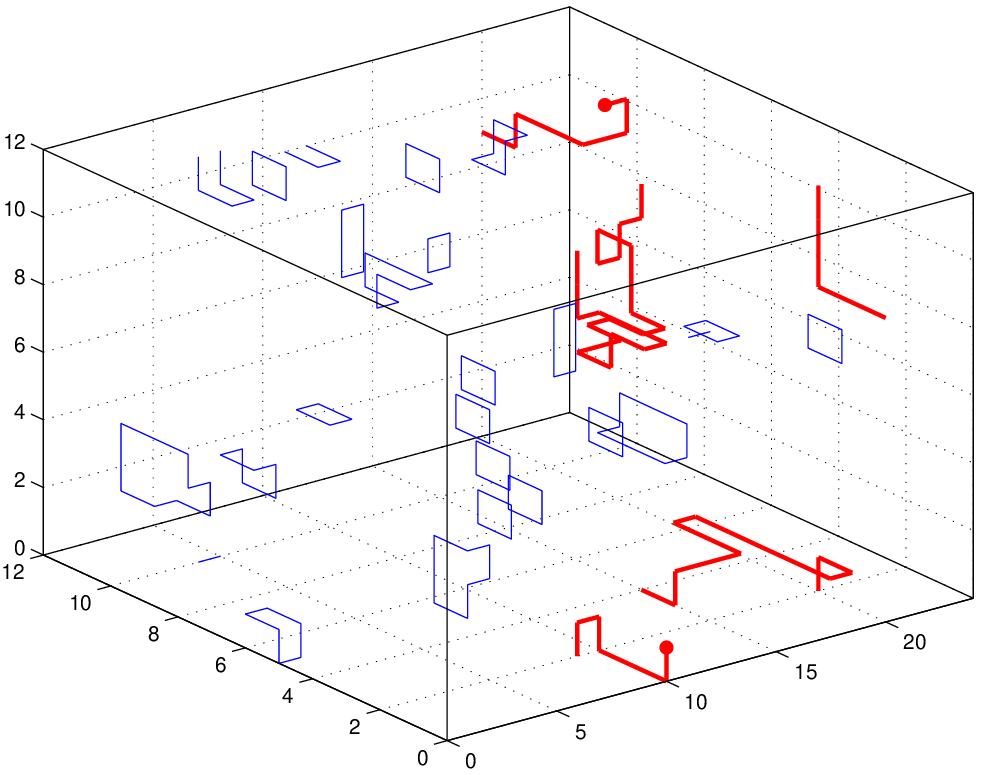,width=0.6\textwidth}
  \caption{A typical configuration on $24 \times 12^2$ with correlation length
  2 ($m = 0.6487 \ldots)$.\label{D3conf}}
\end{figure}

\subsection{$D = 3$, sign problem}

It is trivial to adapt the code from two to three dimensions. On small
lattices $T, L = 4, 6$ we performed similar validation tests as before with
completely accurate and satisfactory results. Note that the formulae for $k_S,
k_V$ in appendix \ref{appA} are equally valid for $D = 2, 3$. It turns out,
however, that now very suddenly as the volume is increased or the mass is
lowered the sign fluctuations abruptly become so violent that no signal is
left in (\ref{kGammara}) and also for ratios of correlations as in
(\ref{meff}) all estimates yield `$0 / 0$' within errors: the sign problem.
For large enough mass loops remain small and predominantly planar. Such loops
are as in two dimensions with phase one. For a demonstration we show in table
\ref{Res3D} results for two cases just before trouble strikes. In figure
\ref{D3conf} the last configuration of our run at $\omega L = 6$ is shown. For
$\omega L = 4$ and the same lattice size no meaningful results can be obtained
anymore. Integrated autocorrelations times were close to $1 / 2$ for all the
quantities studied. Although the observables in the dimer ensemble are
complex, the averages of the imaginary parts vanish within errors as they have
to, since they are parity odd. This was first checked and then used before
forming quotients. A bias was not used here, $\rho \equiv 1$.

\begin{table}[htb]
\centering
  \begin{tabular}{|c|c|c|c|c|}
    \hline
    $\omega L$ & {\large $\frac{\langle \langle \delta_{u, v} \Phi^{(0, 0, 0)} \rangle
    \rangle}{\langle \langle \delta_{u, v} \rangle \rangle}$} &{\large $\frac{\langle
    \langle \delta_{u, v} \Phi^{(1, 0, 0)} \rangle \rangle}{\langle \langle
    \delta_{u, v} \rangle \rangle}$} & $m_{\tmop{eff}} L (3.5)$ & $M /
    \omega$\\
    \hline
    6 & 0.1344(5) & 0.1343(5) & 6.04(5) & 0.51\\
    \hline
    5 & 0.0168(6) & 0.0164(6) & 4.2(10) & 0.27\\
    \hline
  \end{tabular}
  \caption{Simulation results from $10^7$ iterations on $24 \times 12^2$
  lattices.\label{Res3D}}
\end{table}

In the last column we report a mass $M$ which was extracted from the time
slice correlation $\langle \langle \delta_{u_0 - v_0, t} \rangle \rangle$
without phase factors and using the time periodic $\delta$-function. It also
shows a mass-plateau and, as we see, it is smaller than the physical mass. Due
to interference effects the spinor correlation decays faster than this
`geometric' one. From this observation one could think that the loops and
strings have the `wrong' size. We tried to generate them with a modified mass
parameter $m + \Delta m$ such that $M \approx \omega$ is achieved and then
reweighted the observables to the true mass, which is easy if the total
monomer number is available. We find however that this simple idea does not
improve the sign problem.

\section{Conclusions}

We have formulated the `worm' algorithm of Prokof'ev and Svistunov for lattice
fermions of the Wilson type with Wilson parameter $r = 1$. As in the Ising
model it estimates stochastically by the Monte Carlo method the untruncated
hopping parameter expansion of the partition function together with the graphs
needed for the full two-point function, which can thus be computed. The PS
algorithm very naturally lends itself to easily keep track of all phase
factors and spin matrices that appear in the expansion. In two space time
dimensions the contributions of all graphs are positive up to finite size
effects and simulations are similarly efficient as in the Ising model. The all
order hopping expansion is also worked out and numerically tested in three
dimensions. Here the weights of fermion loops acquire complex phase factors
and, for small mass and/or large volume lead to numerically uncontrollable
fluctuations. The very sharp borderline was found around correlation length
two for a $24 \times 12^2$ lattice. In particular, the continuum limit cannot
be approached. Clearly here the method has to be complemented for instance by
an improved estimator which sums some part of the contributions analytically
as in cluster methods. No such method is known at present for the system at
hand.

For two dimensional fermions we now plan to add the interaction of the
Gross-Neveu model. For the O($N$) invariant model the Majorana fermion
discussed here then has to be replicated $N$-fold. For each multi-dimer
configuration there are now between $K (x) = 0$ and $K (x) = N$ monomers at each
site. The four fermion interaction can be enforced by integrating over the
common external field $m (x)$ site by site with the appropriate Gaussian
weight yielding a $K (x)$-dependent total weight as already discussed in
{\cite{Wolff:2007ip}}. In this way a coupling between the $N$ `flavors'
arises. The worm head and tail $u$ and $v$ now refer to one of the flavors.
When $u = v$ is reached with random re-location also a new flavor-index is
chosen randomly. Further details still have to be worked out.

\tmtextbf{Acknowledgments}. I would like to thank Oliver B\"ar, Rainer Sommer and Willi Rath
for discussions. Part of this work was carried out during a one month visit to
UCSD (San Diego) and I wish to thank Julius Kuti and the whole high energy
physics group for making my stay a most pleasant experience. I would like to
thank the Deutsche Forschungsgemeinschaft for support in the framework of SFB
Transregio 9.

\appendix\section{Timeslice correlations of free fermions\label{appA}}

We prefer to re-introduce the lattice spacing $a$ for this appendix.

For our numerical test we need to know the free Wilson propagator (\ref{Geq})
in the time-momentum basis
\begin{equation}
  \breve{G} (x_0, p_k) = a^{D - 1} \sum_{x_k} G (x) \mathe^{- i p_k x_k}
\end{equation}
where Latin vector indices $k = 1, \ldots, D - 1$ refer to spatial components
only. Spatial momenta are quantized as demanded by the spatial boundary
conditions $\varepsilon_k$ and sizes $L_k$. From the Fourier expansion of $G$
it is easy to see that $\breve{G}$ is given in terms of a function $f (x_0)$
by
\begin{equation}
  \breve{G} (t, p_k) = (- i a \mathring{p}_k \gamma_k + 1 + a M) f (t) -
  \frac{1}{2} (1 + \gamma_0) f (t + 1) - \frac{1}{2} (1 - \gamma_0) f (t - 1)
\end{equation}
with $a \mathring{p}_{\mu} = \sin (a p_{\mu})$, $M = m + \frac{a}{2} \hat{p}_k
\hat{p}_k$, $a \hat{p}_{\mu} = 2 \sin (a p_{\mu} / 2)$. The function itself
reads
\begin{equation}
  f (t) = \frac{1}{a T} \sum_{p_0}  \frac{\mathe^{i p_0 t}}{\mathring{p}_0^2 +
  \mathring{p}_k \mathring{p}_k + (M + \frac{a}{2} \hat{p}_0^2)^2} .
\end{equation}
We find it useful to first evaluate $f$ in the limit of infinite time extent
$T = L_0 \rightarrow \infty$. We demand $1 + a M > 0$ and introduce
\begin{equation}
  \sinh (a \omega / 2) = \frac{a}{2} \left[ \frac{M^2 + \mathring{p}_k \mathring{p}_k}{1
  + a M} \right]^{1 / 2}, \hspace{1em} a \omega \geqslant 0.
\end{equation}
With $z = \mathe^{i a p_0}$ we now arrive at the contour integral
\begin{equation}
  f_{\infty} (t) = \frac{- 1}{1 + a M} \oint \frac{d z}{2 \pi i} \frac{z^t}{(z
  - \mathe^{a \omega}) (z - \mathe^{- a \omega})}
\end{equation}
and obtain
\begin{equation}
  f_{\infty} (t) = \frac{1}{2 (1 + a M) \sinh a \omega}  \mathe^{- \omega |t|}
  .
\end{equation}
The solution that is (anti)periodic over finite time now follows from
\begin{equation}
  f (t) = \sum_{n = - \infty}^{\infty} f_{\infty} (t + n T) (- 1)^{n
  \varepsilon_0}
\end{equation}
and reads explicitly for $0 \leqslant t < T \text{}$
\begin{equation}
  f (t) = \frac{1}{2 (1 + a M) \sinh a \omega} \times \left\{
  \begin{array}{lll}
    \cosh [\omega (T / 2 - t)] / \sinh (\omega T / 2) & \tmop{for} &
    \varepsilon_0 = 0\\
    \sinh [\omega (T / 2 - t)] / \cosh (\omega T / 2) & \tmop{for} &
    \varepsilon_0 = 1
  \end{array} \right. .
\end{equation}
For $0 < t < T$ this yields
\begin{equation}
  \tmop{tr} [ \breve{G} (x_0, p_k)] = \frac{M - a \hat{\omega}^2 / 2}{(1 + a
  M) \mathring{\omega}} \times \left\{ \begin{array}{lll}
    \cosh [\omega (T / 2 - t)] / \sinh (\omega T / 2) & \tmop{for} &
    \varepsilon_0 = 0\\
    \sinh [\omega (T / 2 - t)] / \cosh (\omega T / 2) & \tmop{for} &
    \varepsilon_0 = 1
  \end{array} \right. \label{GS}
\end{equation}

with $a \hat{\omega} = 2 \sinh (a \omega / 2)$, $a \mathring{\omega} = \sinh (a
\omega)$, and
\begin{equation}
  \tmop{tr} [\gamma_0  \breve{G} (x_0, p_k)] = \frac{1}{1 + a M} \times
  \left\{ \begin{array}{lll}
    \sinh [\omega (T / 2 - t)] / \sinh (\omega T / 2) & \tmop{for} &
    \varepsilon_0 = 0\\
    \cosh [\omega (T / 2 - t)] / \cosh (\omega T / 2) & \tmop{for} &
    \varepsilon_0 = 1
  \end{array} \right. . \label{GV}
\end{equation}
At zero time separation $\tmop{tr} [\gamma_0  \breve{G} (0, p_k)] = 0$
vanishes and
\begin{equation}
  \tmop{tr} [ \breve{G} (0, p_k)] = \frac{1}{1 + a M} + \frac{M - a
  \hat{\omega}^2 / 2}{(1 + a M) \mathring{\omega}} \times \left\{
  \begin{array}{lll}
    \coth [\omega T / 2] & \tmop{for} & \varepsilon_0 = 0\\
    \tanh [\omega T / 2] & \tmop{for} & \varepsilon_0 = 1
  \end{array} \right.
\end{equation}
Exact results for the zero momentum (for $\varepsilon_k = 0$) correlations
\begin{equation}
  k_S (x_0) = - a^{D-1} \sum_{x_k} \langle \overline{\xi} (0) \xi (x) \rangle =
  \tmop{tr} [ \breve{G} (x_0, 0)]
\end{equation}
and
\begin{equation}
  k_V (x_0) = - a^{D-1} \sum_{x_k} \langle \overline{\xi} (0) \gamma_0 \xi (x)
  \rangle = \tmop{tr} [\gamma_0  \breve{G} (x_0, 0)]
\end{equation}
follow from the above formulas.

Incidentally these free field or tree level results form an example how
Symanzik improvement for Wilson fermions works. If we multiply the left hand
sides of (\ref{GS}) and (\ref{GV}) by the wavefunction improvement factor $1 +
a m$ and eliminate on the right hand side $m$ by the pole mass
$m_{\tmop{pole}} = a^{- 1} \ln (1 + a m)$, then the finite size continuum
limit ($T / a \rightarrow \infty$ at fixed $t / T, m_{\tmop{pole}} T, p_k T$)
is reached at a rate proportional to $(a / T)^2$. For $p_k = 0$, where $\omega
= m_{\tmop{pole}}$ holds, lattice artifacts are even completely eliminated.

In the correlations above one may also safely take the limit $m \rightarrow 0$
at finite $T$ for {\tmem{antiperiodic}} boundary conditions $\varepsilon_0 =
1$. This setup may be interpreted as a finite temperature which supplies an
infrared regulator, similar as but simpler than in the Schr\"odinger
functional {\cite{Luscher:1992an}}. In this limit $k_V$ becomes constant while
$k_S$ vanishes. In $D = 2$ the latter may be viewed as a consequence of the
discrete chiral symmetry $\xi \rightarrow i \gamma_0 \gamma_1 \xi$ under which
$k_S$ is odd. On the lattice with Wilson fermions this symmetry must emerge in
the continuum limit at the critical mass, zero in the free case. Due to the
(accidental) complete improvement it here holds already on the lattice at
{\tmem{all}} nonzero separations.

As discussed in {\cite{Korzec:2006hy}} $k_S, k_V$ may be used to formulate
massless nonperturbative renormalization conditions in the interacting 
Gross-Neveu model. The field can be normalized by keeping for instance $k_V (T / 2)
= 1$ while the critical mass may be determined by tuning to $k_S (T / 4) = 0$.
Note that $k_S (T / 2)$ vanishes due to time reflection invariance combined
with antiperiodicity for all mass values and is not suitable to fix the chiral
point.

\section{Spin phase: continuum and lattice\label{appB}}

We now consider an alternative way to compute the loop phase (\ref{wlam})
\begin{equation}
  w (\lambda) = - \tmop{tr} [P (m_1) P (m_2) \cdots P (m_{| \lambda |})]
\end{equation}
from the sequence of lattice unit vectors $m_i$. An earlier discussion of this
quantity is found in {\cite{Stamatescu:1980br}}. As a technical simplification
we assume for a while that straight sections are absent, $m_{i + 1} \cdot m_i
= 0$. We can rotate one projector into the next, for example
\begin{equation}
  P (m_1) P (m_2) = R (m_1, m_2) P (m_2) R (m_1, m_2)^{- 1} P (m_2) =
  \frac{1}{\sqrt{2}} R (m_1, m_2) P (m_2) \label{Preduce}
\end{equation}
with the $\pi / 2$ spinor rotation matrix

\begin{equation}
  R (m_1, m_2) = \mathe^{\frac{\pi}{8} [ \Sh{m_1} \;\; , \Sh{m_2} \;\;]}
  \label{Cuberot}
\end{equation}
in the usual notation $\Sh{m} = m_{\mu} \gamma_{\mu}$. By iterating this step
we arrive at
\begin{equation}
  - w (\lambda) = 2^{- (| \lambda \vdash 1) / 2} \tmop{tr} [R_{\tmop{loop}} R
  (m_1, m_{| \lambda |}) P (m_{| \lambda |})]
\end{equation}
with
\begin{equation}
  R_{\tmop{loop}} = R (m_1, m_2) R (m_2, m_3) \cdots R (m_{| \lambda | - 1},
  m_{| \lambda |}) R (m_{| \lambda |}, m_1) .
\end{equation}
As the loop closes this total rotation must bring $m_1$ back to itself, it
belongs to the stability group of this direction.

The rotations introduced in (\ref{Cuberot}) can be characterized by mapping a
cube onto itself by rotations by $\pi / 2$ through any lattice plane. Products
of such operations do not preserve the form (\ref{Cuberot}) but are general
members of the cubic group. The number of elements is given by twice the
number of orientations of a $D$-dimensional cube along the lattice axes
(without reflections) that is $2^D D!$ in $D$ dimensions. The twofold
multiplicity is because $\pm R$ in the spinor representation corresponds to
only one re-orientation of the cube as usual with spinor rotations.

In $D = 2$ the cubic group is abelian and has 8 elements. Here
$R_{\tmop{loop}}$ adds up the half-angles of the rotations that a
test-particle taken around the loop experiences. For a non-intersecting closed
loop we get $R_{\tmop{loop}} = - 1$ for a $2 \pi$ rotation. Then we find in
total
\begin{equation}
  w (\lambda) = (-)^{\nu + 1} 2^{- C (\lambda) / 2}
\end{equation}
with the number of complete rotations $\nu$ and the number of corners along
the loop $C (\lambda)$. This form of the result now also covers the case where
$m_{i + 1} = m_i$ can occur, a straight section not counted in $C (\lambda)$.

In $D = 3$ the cubic group is non-abelian and has 48 elements. Now
$R_{\tmop{loop}}$ can only be a rotation around the $m_1$ axis,
\begin{equation}
  R_{\tmop{loop}} = \pm \mathe^{i \frac{\alpha}{2}  \Sh{m_1}}
\end{equation}
with $\alpha$ a multiple of $\pi / 2$. Then one easily evaluates
\begin{equation}
  \tmop{tr} [R_{\tmop{loop}} R (m_1, m_{| \lambda |}) P (m_{| \lambda |})] =
  \frac{1}{\sqrt{2}} \mathe^{- i \frac{\alpha}{2}} .
\end{equation}
The sign ambiguity of the spinor transformation (2 to 1 covering) has been
absorbed into the $\alpha$ and the phase is in Z(8). For $m_1 = m_{| \lambda
|}$ again the factor $1 / \sqrt{2}$ has to be dropped.

It is interesting to note that a strictly analogous construction can be made
in the continuum. Let us assume a closed curve $\vec{\gamma} (t), t \in [0,
1]$. For a regular parametrization we have $\dot{\vec{\gamma}} \not= 0$ and
form $m (t) = \text{$\dot{\vec{\gamma}}$} / | \text{$\dot{\vec{\gamma}}$} |$,
a mapping from $S_1$ to $S_{D - 1}$. We now move $m (t)$ along the loop by
infinitesimal rotations in spinor form
\begin{equation}
  \dot{\Sh{m}} = \dot{\Sh{m}}\,  \Sh{m}^2 = \frac{1}{2} [ \dot{\Sh{m}} \,
  \Sh{m}, \Sh{m}]
\end{equation}
using $\dot{m} \cdot m = 0$. The solution of
\begin{equation}
  \frac{d}{d t} R (t) = \frac{1}{2}  \dot{\Sh{m}}\,  \Sh{m} R (t),
  \hspace{1em} R (0) = 1
\end{equation}
is formally given by the path-ordered integral
\begin{equation}
  R (t) =\mathcal{P} \exp \left\{ \frac{1}{2} \int_0^t d \tau \dot{\Sh{m}}
  (\tau) \Sh{m} (\tau) \right\} .
\end{equation}
The direction $m$ is parallel transported according to
\begin{equation}
  \Sh{m} (t) = R (t) \Sh{m} (0) R^{- 1} (t) .
\end{equation}
At $t = 1$ we must recover $\Sh{m} (1) = \Sh{m} (0)$ and thus $R (1)$ must
induce a rotation that has $m (0)$ as a fixed point. In $D = 2$ there is no
nontrivial such rotation, but since both $R = 1$ and $R = - 1$ induce the
trivial rotation of $m (0)$, it is possible to find $R (1) = - 1$. Indeed the
latter result emerges for a loop with an odd number of windings. In $D = 3$ $R
(1)$ can induce an arbitrary rotation around the $m (0)$ axis, $R (1) = \exp
[i \alpha \Sh{m} (0) / 2]$. Thus there is a phase, now a general element of
U(1), associated with a closed loop. It does not depend on where the loop is
cut by the parametrization: going to another point, $R (1)$ just gets
similarity-transformed. It is a property just of the loop itself. It seems
very likely that this is a known concept in differential geometry, however no
reference is known to the author.

\end{document}